# The Kilodegree Extremely Little Telescope: Searching for Transiting Exoplanets in the Northern and Southern Sky

Jack Soutter[1], Jonti Horner[1,2], Joshua Pepper[3,4] and the KELT Science Team

[1] University of Southern Queensland West St, Toowoomba QLD 4350, Australia
[2] Australian Centre for Astrobiology, UNSW Australia, Sydney, NSW 2052, Australia
[3] Department of Physics and Astronomy, Vanderbilt University, Nashville, TN 37235, USA
[4] Department of Physics, Lehigh University, Bethlehem, PA 18015, USA

**Summary:** The Kilodegree Extremely Little Telescope (KELT) survey is a ground-based program designed to search for transiting exoplanets orbiting relatively bright stars. To achieve this, the KELT Science Team operates two planet search facilities - KELT-North, at Winer Observatory, Arizona, and KELT-South, at the South African Astronomical Observatory. The telescopes used at these observatories have particularly wide fields of view, allowing KELT to study a large number of potential exoplanet host stars. One of the major advantages of targeting bright stars is that the exoplanet candidates detected can be easily followed up by small, ground-based observatories distributed around the world. This paper will provide a brief overview of the KELT-North and KELT-South surveys, the follow-up observations performed by the KELT Follow-up Collaboration, and the exoplanet discoveries confirmed thus far, before concluding with a brief discussion of the future for the KELT program.

## Introduction

Since the discovery of the first exoplanets, over two decades ago (e.g. [1][2][3]) the field of extrasolar planetary science has experienced rapid growth. That growth has been driven by two main factors. The first is the ever-lengthening temporal baseline over which observations have been carried out, which is allowing the discovery of ever more distant planets, primarily through the Radial Velocity method[1], including the detection of the first 'Jupiter Analogues' around other stars (e.g. [4][5][6]). The second factor is the development of new technologies, tools, and programs that allow the study of an ever-greater number of stars in a variety of different manners (e.g. [7][8][9]). The influence of these new programs is perhaps best illustrated by the success of the Kepler mission [10][11][12], which has, to date, discovered 1030 planets[2].

When it comes to exoplanet surveys, two techniques greatly dominate the catalogue of discovered planets: Radial-Velocity (RV) and transit photometry. RV observations look for the subtle variations in a star's spectrum caused by the gravitational pull of orbiting bodies [13]. In contrast, the transit photometry method instead analyses a star looking for a drop of its visual brightness, caused by an exoplanet transiting in front of the star's disc [8]. While RV surveys have resulted in a large number of discoveries, they remain limited to the brightest stars in the night sky, with current technological limitations of stellar spectroscopy on the telescopes involved typically preventing RV surveys from observing many stars with an apparent visual magnitude fainter than ~8 without prohibitively long observations. Surveys searching for transiting exoplanets, on the other hand, only require photometric observations of stars, and as such are well suited for ground-based surveys that simultaneously target large numbers of fainter stars[3] [14].

---
[1] The discussion of the various techniques used to discover planets around other stars is beyond the scope of this work – but we direct the interested reader to [13] for more details.
[2] The tally of planets discovered by the Kepler spacecraft continues to rise rapidly – the value of 1030 confirmed planets was obtained from the Kepler website, http://kepler.nasa.gov, on 18th November 2015, at which time Kepler had a further 4696 candidate planets awaiting confirmation.

For this reason, recent years have seen a proliferation in wide-field exoplanet transit surveys. In addition to KELT, the HATNet [15] and SuperWASP [16] programs have been running for several years, and have made a number of interesting discoveries in both the northern and southern skies [17][18]. The KELT survey, which operates in both hemispheres, typically targets stars brighter than those observed by other transit observation programs. To do this, the KELT surveys use two small, wide-field telescopes located in the southern and northern hemispheres to provide a rich supply of possible exoplanet hosting stars within the range of $8 < V < 12$ mag [19][20].

One of the great benefits of targeting bright stars with a transit survey such as KELT is that any planets found can then be followed up using other small astronomical facilities, allowing their accurate characterisation. To this end, the KELT science team incorporates a network of follow-up facilities located around the world, to facilitate timely and detailed follow-up observations.

The process of following up KELT discoveries typically involves two discrete steps. The first is confirmation follow-up. Here, the observed dips in a star's brightness due to the potential transit of a planet are used to create an ephemeris, and thereby predict the timings of future transits. The telescopes in the follow-up network can then target the potential planet-hosting star at the right time, and hopefully confirm that it dims as expected. This confirmation process takes place before the potential existence of the planet is widely publicised, and allows the KELT team to better characterise the orbit of the planet prior to its announcement.

Once the target has been confirmed as an exoplanet candidate, the true benefit of the brightness of the host star comes into play – with a bright host, a plethora of additional characterisation observations are possible. Observing future transits with larger telescopes allows the diameter of the newly discovered planet to be determined, and since the stars are bright, it is possible to also obtain measurements of the star's radial velocity variations – which allow the mass of the planet to be determined. By combining these observations, then, we are able to calculate the density of the planet in question, allowing us to determine whether it is most likely rocky or gaseous. Beyond this, further follow-up observations can allow the nature of the planet to be studied in still more detail. For example, measurements of the planet host star's radial velocity *during a transit* enable the measurement of the Rossiter-McLaughlin effect [21][22], allowing the determination of the orientation of the planet's orbit with respect to the spin axis of its host star. If sufficiently detailed spectra of the star can be obtained during either primary or secondary transit, the broad atmospheric composition of the planet can be determined [23][24].

---

[3] It is worth noting that while photometric data is perfect for analysing exoplanet candidature, it alone can not be used for confirmation. Additional radial velocity measurements are required for confirmation and characterisation. This is the reason that many of the Kepler spacecraft's detections are listed as exoplanet candidates rather than confirmed exoplanets.

In the following section, we provide a basic overview of the KELT survey, before moving on to a discussion of the follow-up process and partners. We then summarise some of KELT's key discoveries to date, before concluding with a short discussion of the future goals of the KELT program.

# The KELT survey

The KELT program was launched in 2005, with the installation of the KELT-North telescope at Winer Observatory, Arizona. After that telescope had been successfully operating for several years, the KELT-South telescope was installed, at the South African Astronomical Observatory in Sutherland, South Africa. Both telescopes have operated almost continually since their installation, whenever the weather was suitable, and have allowed a large database of potential exoplanet candidates to be assembled. In order to follow-up on the objects in the catalogue, a significant number of other observatories have been engaged in the KELT program, as described below.

*Optical assembly*

The KELT project uses two similar optical assemblies in the northern and southern hemisphere to perform its wide-field survey. The optical assembly of each survey instrument consists of a CCD detector and a wide-field camera lens mounted on a robotic telescope mount, as shown in Figure 1.

Both the KELT-North and KELT-South detectors are equipped with a Mamyia 645 80mm f/1.9 medium-format manual focus lens with a 42mm aperture, which provides a 26° x 26° field of view, which corresponds to an image scale of approximately 23" pixel$^{-1}$ [19][20]. KELT-North and KELT-South also use similar CCD detectors; KELT-North uses an Apogee AP16E with 4096 x 4096 9μm pixels, while KELT-South operates using an Apogee U16M with 4096 x 4096 9μm pixels (both models use Kodak KAF-16801 front-illuminated CCDs). Both cameras are mounted on German-equatorial Paramount ME Robotic Telescope mounts.

*KELT Locations*

The KELT telescopes are located in both the southern and northern hemispheres to allow for complete sky coverage over the course of the year. The KELT-North telescope is housed at the Irvin M. Winer Memorial Mobile Observatory in Sonoita, Arizona. The site is located at N 31°39'53", W 110°36'03" at an elevation of 1515m. The KELT-South telescope is located in Sutherland, South Africa at the South African Astronomical Observatory (32°22'46" S, 20°38'48" E, elevation 1760m). Both locations yield relatively consistent clear skies, yielding an observing up-time of approximately 60% and 70%, respectively. The locations of the two KELT telescopes can be seen in Figure 2, below.

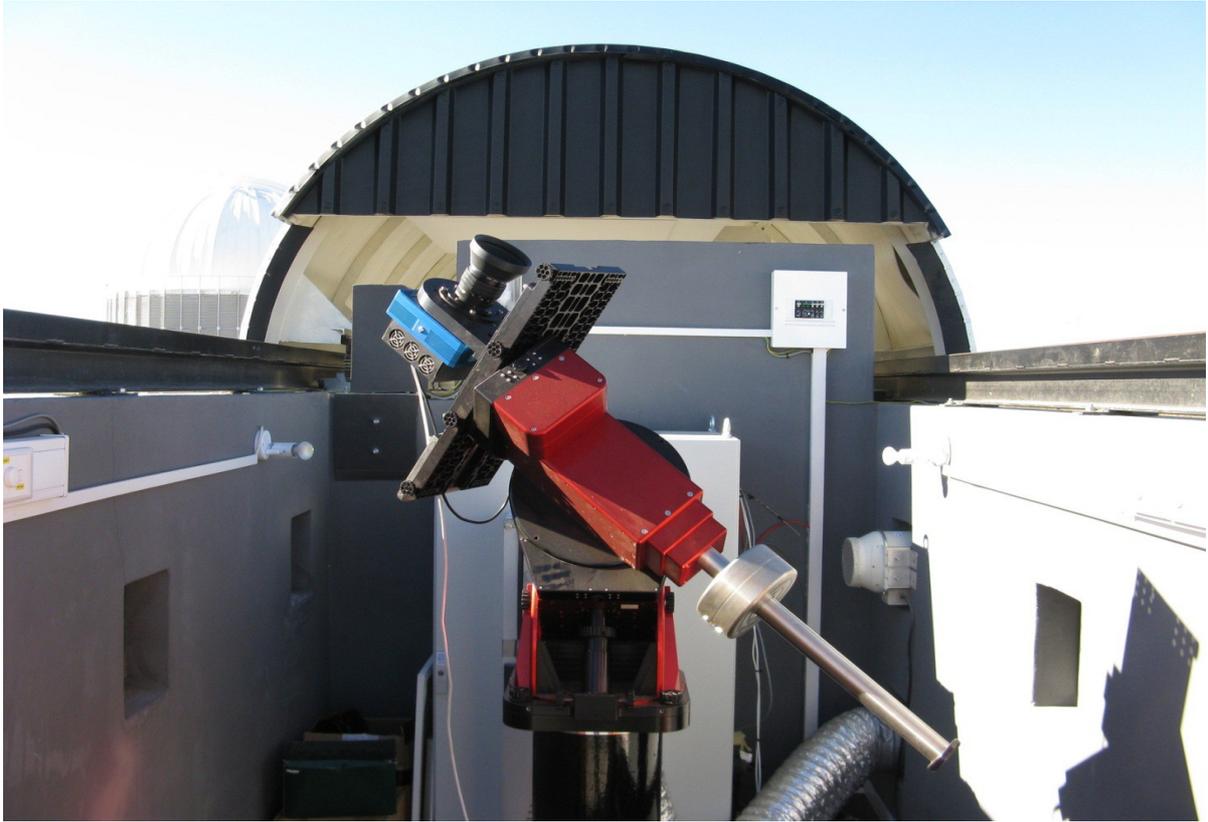

*Figure 1: The KELT-South telescope located at the South African Astronomical Observatory in Sutherland, South Africa. The telescope is fully automated, and designed such that the dome opens and observations proceed only when conditions are appropriate. (Image courtesy of Dr Joshua Pepper, Vanderbilt University)*

*Observations and Data Reduction*

Both KELT telescopes are designed to operate robotically on any night when the weather is suitable; with winds not exceeding the loading limit of the instrument, no heavy cloud cover, etc. If the conditions are suitable, the roof will open and the telescope will observe. It will observe a number of pre-selected fields across the sky, firstobserving all visible fields east of the local meridian, and then preforming a meridian flip and continuing the process to the west of the meridian. That process is repeated until dawn.

After the images are taken, the data is transferred online back to Ohio State (for KELT-North), and Vanderbilt University (for KELT-South). The data is then processed so that it can be added to a combined KELT candidate database, incorporating data and candidates from both telescopes.

Any images of low quality are removed based on their high background sky count, which is typically caused by excessive moonlight or cloud cover. After the compromised data has been removed, basic data reduction is performed through the subtraction of dark and bias images from the flat field images and raw data, a process that is followed by flat fielding. The final two automated steps involve the use of the astrometry software package from astrometry.net [25], which overlays a world coordinate system (WCS) into the .fits data files. Once the images have been properly calibrated, the final step is to generate light-curves for each star.

This is accomplished by performing image subtractions using a heavy modified version of the ISIS imaging package [26]. The ISIS package performs spatial recognition and aligns each image to a reference before creating a differenced image. These images allow for precise flux measurements and the creation of accurate light-curves. This information is then added to the KELT database to be accessed by the follow-up team.

## Follow-Up Partners

As stated above, the KELT survey is designed to detect giant planets in orbits close to their host stars, and to target main-sequence stars with an apparent visual magnitude between 8 and 10. This has two major upsides.

First, it allows for exoplanet detection around stars that are fainter than those typically targeted by Radial Velocity planet search programs. Fainter stars require significantly longer exposures, and, fainter than ninth magnitude, such exposures become prohibitive when attempting to carry out a large survey. However, should planets be found around stars in this magnitude range (eighth to tenth), radial velocity follow-up observations are relatively straightforward. When the requirement to observe many stars many times is removed, resources can be efficiently allocated to observe the single target over longer periods, once there is evidence of an existing exoplanet. This means that the planets detected by KELT are ideally placed for the perfect combination of observational techniques – if a planet can be observed by both radial velocity and transit observations, a far more complete characterisation is possible than for planets observed by only one method.

Secondly, the relative brightness of the KELT target stars allows for timely photometric follow-up with small ground-based telescopes, and KELT can therefore take advantage of small observatories hosted by their partner institutions. In order to best make use of this ever-growing catalogue of planet candidates, and achieve timely confirmation and publication of their results, the KELT team have built a large collaborative network with partner institutions that operate their own observatories scattered around the planet. This collection of collaborators makes up the wider KELT Follow-up Team. There are currently over 30 observatories across both hemispheres that make up that team, and provide the KELT project with photometric and (where appropriate) RV follow-up observations.

There are typically hundreds of exoplanet candidates in the KELT database at any one given time, meaning that any observatory in the KELT network will have at least 2-3 viable targets every night. This density of targets allows for constant follow-up observations to be carried out year round, allowing for the removal of false-positives from the database and to refine the transit timing of any legitimate exoplanets. This widely distributed approach is already yielding great results, with a number of planet discoveries already published, and several more in preparation.

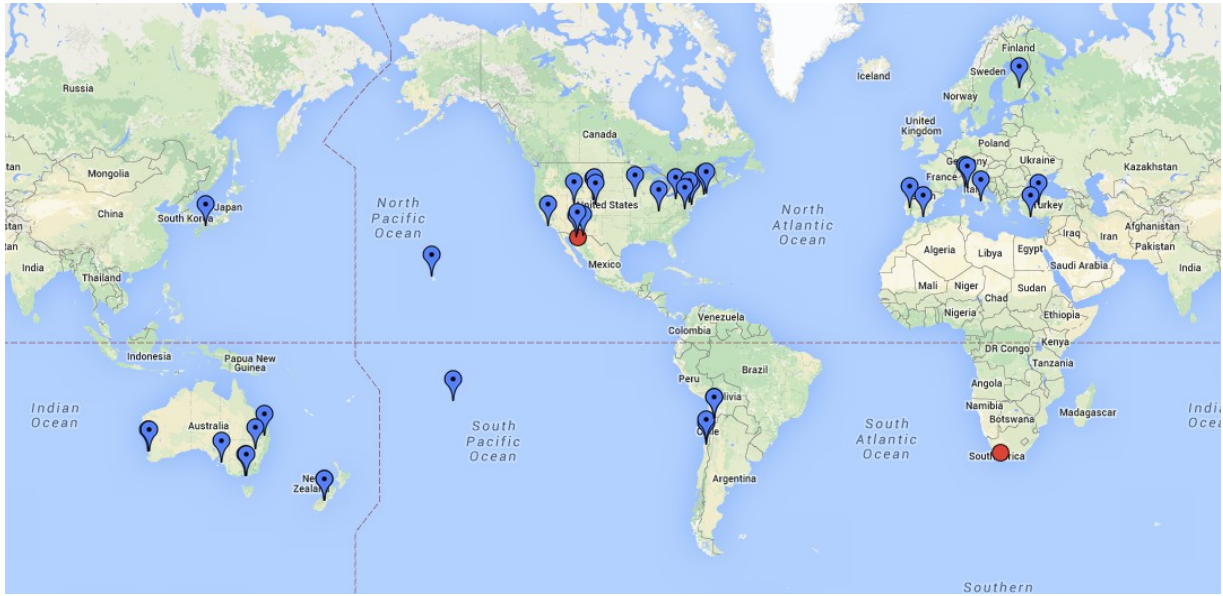

*Figure 2: Locations of the KELT-North and KELT-South telescopes (marked in red), and the various observatories around the world that provide KELT with follow-up photometry (marked in blue). Images courtesy of Joshua Pepper, Vanderbilt University and Google Maps.*

## Discoveries

The KELT survey has already discovered and published a number of exoplanets. Here, we present a concise overview of some of KELT's discoveries to date.

*KELT-1b*

The first planet-sized object to be discovered by the KELT project, KELT-1b is a highly irradiated, transiting brown dwarf[4] orbiting its companion at a semi-major axis of only 0.02466±0.00016 au. Following the detection of the transits of KELT-1b, a program of spectroscopic, radial velocity and photometric follow-up was carried out by the KELT team, which resulted in a detailed characterisation of the object. As a result of that follow-up work, we know that KELT-1b weighs in at ~27$M_J$, placing it firmly within the brown-dwarf mass range. The follow-up observations also revealed that the orbit of KELT-1b is not significantly inclined to its host's equator, being consistent with zero obliquity, and suggest that the parent star is tidally locked to the orbit of KELT-1b. Finally, the follow-up observations also revealed the possible presence of an M-dwarf companion to the primary star, located at a separation of 588 milliarcseconds. Full details of the discovery, and the follow-up work carried out, can be found in [27].

---

[4] Brown dwarfs are objects that tread the line between stars and planets. The dividing line between an object being considered a planet and a brown dwarf lies at the Deuterium burning limit, at roughly 13 times the mass of Jupiter. In theory, the temperature and pressure in the core of such an object should be high enough for them to undergo a short period of Deuterium fusion, without ever reaching the conditions required to fuse Hydrogen. The dividing line between brown dwarfs and stars lies at the Hydrogen burning limit, at ~80 times Jupiter's mass. The official classification can be found at [28] and [29].

*KELT-2Ab*

KELT-2Ab is a hot Jupiter (~1.52$M_J$) in orbit around the primary star of the HD 42176 binary system, a late F-type star. At the time of its discovery, the mass and radius of KELT-2Ab placed it in a unique position amongst planets discovered orbiting bright stars. Indeed, it was the only known transiting planet in the mass range 1.2$M_J$ and 3$M_J$ orbiting such a bright star (V≈ 8.7). In many other ways, however, the planet is fairly unremarkable, being a fairly typical hot Jupiter (orbital period of just over four days, and slightly more massive and larger than Jupiter). During the follow-up observations of KELT-2Ab, the science team was also able to demonstrate for the first time that KELT-2A and KELT-2B were gravitationally bound to each other as a binary system, which in turn allowed them to infer that KELT-2B, the secondary star in the system, must be an early K-dwarf. Full details of the discovery and follow-up of KELT-2B can be found in [30].

*KELT-3b*

The third confirmed exoplanet from the KELT survey was KELT-3b ([31]), an inflated hot Jupiter with a mass of ~1.48$M_J$ and a semi-major axis of ~0.041 au. The host star itself is a late F-type star with a visual magnitude of 9.8. While KELT-3b is a typical Hot Jupiter there is an interesting comparison between its host star and the planet-hosting star of HAT-P-2. The two stars have near identical masses, radii, metallicities, temperatures and ages. However, the two stars display vastly different rotational velocities, with KELT-3 rotating approximately half as quickly as HAT-P-2.

This difference in the rotational velocity of two stars that are otherwise almost identical may well have something to do with the nature of their planets. HAT-P-2b is significantly more massive than KELT-3b, and so the faster spin rate of HAT-P-2 (which is still longer than the orbital period of its planet) might feasibly be related to tidal star-planet interaction, particularly since the mass of the companion is likely comparable to the mass of the host's convective envelope [32]. In addition, like KELT-2Ab, there are suggestions that KELT-3 might well have a common proper motion companion – although its binary has not yet been definitively confirmed. Given that one of the proposed mechanisms by which hot-Jupiters could be created involves distant perturbations from massive companion objects, and evolution through a combination of Kozai-resonant oscillation and tidal-damping, the presence of companions in hot-Jupiter systems is always an interesting observation.

*KELT-4Ab*

KELT-4Ab is an inflated hot-Jupiter that orbits the brightest component of a hierarchical triple stellar system ([33]). In other words, KELT-4Ab orbits around KELT-4A, an F-star that is somewhat more massive and hotter than our Sun. Based on the Washington Double Star Catalog ([34]), KELT-4 had been identified as a common proper motion binary, with the components separated by 1.5" (following observations by [35] in 1973). The KELT follow-up program carried out Adaptive Optics observations of the system using the Keck II telescope in 2012. These revealed that the secondary companion (KELT-4B) was actually itself double, a binary separated by just ~49 mas, which the discoverers estimate to be twin K-stars, with $T_{eff}$ ~ 4300 K.

As such, KELT-4Ab is just the third known transiting planet within such a hierarchical triple system. Due to the nature of the system, it is natural to consider whether Kozai-resonant interactions between KELT-4Ab and the KELT-4BC binary pair could have helped to drive the planet into its current tight orbit about KELT-4A. [33] detail calculations examining this question, and show that such migration is certainly feasible for all eccentricites of the KELT-4A – BC orbit, although there is a sub-section of orbital element space in which such interactions could not engineer sufficient Kozai-resonant effects on the planet's orbit (namely those with low/moderate eccentricites and a near-coplanar planet-binary orbital architecture). Future radial-velocity observations to measure the Rossiter-McLaughlin effect for this system will be of great interest, since any mis-alignment between the rotation of the star and the orbit of its planet could be indicative of such a migration history.

*KELT-6b*

The least massive of all of the KELT discoveries to date, KELT-6b is a hot Saturn (~0.43$M_J$) orbiting around a metal-poor ([Fe/H] = -0.28), V=10.38, late F-type host star ([36]). KELT-6b is very similar to the well-studied exoplanet HD 209458b, other than for the metal-poor nature of its host star. As with the other planets detected by the KELT survey, KELT-6 is among the twenty brightest stars currently confirmed to host exoplanets, making it an excellent target for current and future characterisation work.

*KELT-7b*

KELT-7b is a relatively typical, somewhat inflated, hot-Jupiter (~1.28$M_J$, ~1.53$R_J$; [37]). Of the planets found to date by the KELT program, KELT-7b orbits the brightest star. Its host is significantly more massive and hotter than the Sun (~1.54 $M_\odot$, $T_{eff}$ ~6789 K). Furthermore, the host is a rapid rotated (~73 km s$^{-1}$), which enhances the strength of the Rossiter-McLaughlin effect. As a result, follow-up observations have revealed that the planet's orbit is well aligned to the plane of the star's oblate equator. Given the planet's relatively small orbital distance, the high luminosity and temperature of its host, and the planet's large 'puffed-up' radius, coupled to the brightness of the star as seen from Earth, KELT-7b is an ideal target for future follow-up and characterisation work.

*KELT-8b*

KELT-8b is one of the most inflated known transiting exoplanets ([38]). It orbits its host star (HD 343246) on a slightly eccentric (e ~0.035) orbit with radius ~ 0.0457 au. Despite being somewhat less massive than Jupiter (M ~0.867 $M_J$), the planet is significantly larger than our own giant, with radius ~1.86 $R_J$, giving inferred density of just ~0.167 g cm$^{-3}$!

KELT-8 (HD 343246) is significantly more massive than the Sun (M ~1.211 $M_\odot$), and is also more metal rich ([Fe/H] ~+0.272). It is thought to be mildly evolved, with the result that it is also larger than the Sun (R ~ 1.67 $R_\odot$), a factor that might play some role in the planet's degree of inflation. As with KELT-7b, the planet's large size means that it is an ideal candidate for follow-up spectroscopic observations to study its atmospheric composition.

*KELT-10b*

KELT-10b the latest confirmed exoplanet published by the KELT survey team [39], and the first confirmed exoplanet resulting from the KELT-South project. KELT-10b is a highly inflated, sub-Jupiter mass (~0.69$M_J$) planet with a semi-major axis of ~0.0529au. Its host star is sub-giant, slightly more massive than the Sun, and it seems likely that the planet is doomed. Analysis of the evolution of KELT-10 suggests that the star will devour KELT-10b as it leaves its current sub-giant phase, at some point in the next billion years.

*KELT-14b and KELT-15b*

The newest additions to the KELT family are two hot Jupiters detected by the KELT-South survey, and announced in a submitted paper posted on arXiv in late September [40]. KELT-14b is an independent discovery of WASP-122b, and illustrates the competitive and complementary nature of exoplanet transit surveys. It has an inferred mass slightly greater than that of Jupiter (M ~ 1.196 $M_J$), and is somewhat inflated, with a radius of ~1.52 $R_J$. It orbits a host star that is somewhat more massive and metal rich than the Sun, and is thought to be close to leaving the main sequence. KELT-15b is yet another inflated hot Jupiter, orbiting a G0 star of approximately the same age as the Sun (~4.6 Gyr). That star, once again, is more massive than our own, at ~1.181 solar masses, and is again thought to be close to leaving the main sequence. It is thought that both planets should be observable in emission at their secondary eclipses, which makes them exciting targets for future atmospheric characterisation.

## KELT in the Future

At this time, over twelve hundred transiting exoplanets have been discovered. The Kepler space satellite discovered the bulk of these, with over one thousand confirmed discoveries to date [10]. Ground-based transit surveys like KELT have also been highly productive, contributing over two hundred planets to the total, a number that will rise in coming years. The transiting planets orbiting bright stars that KELT is discovering are especially valuable due to their accessibility to observations that can characterize their dynamics, energetics, and atmospheres.

In 2017, NASA will be launching the TESS satellite [41], which will scan almost the entire sky over a two-year period, carrying out observations with much higher photometric precision than can be achieved by ground-based telescopes. The TESS program is designed to target stars with I-band magnitudes in the range 4 – 13, and will observe those starts across intervals ranging between one month and one year, depending on their ecliptic latitude. As a result, TESS is expected to discover all transiting planets orbiting those stars with periods less than ten days. For stars closer to the ecliptic poles, TESS' completeness will extend to periods of aboutforty days, or more.

TESS will discover most transiting planets orbiting bright stars as seen from Earth.. Nevertheless, KELT will remain useful for a number of other science goals. These include the discovery and characterisation of eclipsing binary and other periodic variable stars. In addition, TESS will mostly be confined to observing relatively short-period behaviour, while KELT already boasts a 10-year baseline of observations. As a result, KELT will be capable of detecting much longer-term variability in its target stars. The on-going ability of KELT to monitor the sky will continue to be helpful for the characterisation of bright transient events, like supernovae [42].

Furthermore, the worldwide follow-up network assembled for KELT is well positioned to support the upcoming TESS mission, to serve a similar purpose in vetting and confirming many of the TESS planet candidates. The next three years will see a transition among this collaboration from a focus on KELT follow-up to TESS follow-up. At the same time, however, a great quantity of KELT data will remain, and continue to be taken, that can be exploited for a variety of science goals.

Looking further into the future, the next generation of space-based exoplanet search and characterisation programs are currently in development, with proposed missions such as CHEOPS [43], WFIRST [44] and TWINKLE [45] being planned for the coming decade. By placing Australia at the forefront of exoplanet discovery and characterisation, dedicated programs such as the KELT-South survey and the recently funded MINERVA-Australis (a partner to the northern hemisphere MINERVA [46]) are laying the ground work for our future participation on the global stage. The planets discovered by those future missions will require follow-up observations to be carried out on a 24 hour basis, and Australia is ideally positioned to give almost unique longitude and latitude coverage of the night sky. As a result, these facilities will be pivotal in future work, and Australian researchers will be heavily involved in the international teams that are developing those projects.

# Acknowledgements

JH is supported by USQ's Strategic Research Fund: the STARWINDS project.